\documentclass[preprint,12pt]{aastex}
\begin{document}
\title{Flat Cosmology with Coupled Matter and Dark Energies}
\author{A.-M.M. Abdel-Rahman$^1$ and Ihab F. Riad$^2$}
\affil{$^1$Division of Basic Sciences,Faculty of Engineering,
University of Khartoum, P.O.Box 321 , Khartoum 11115,
Sudan\email{abdelmalik@yahoo.com} $^2$Department of Physics, Faculty
of Science, University of Khartoum, P.O.Box 321, Khartoum, 11115,
Sudan, and Department of Astronomy,Faculty of Science,University of
Cape Town,Cape Town,South
Africa\email{ifriad@circinus.ast.uct.ac.za}}

\begin{abstract}

Three models of a flat universe of coupled matter and dark
energies with different low-redshift parameterizations of the dark
energy equation of state are considered.  The dark energy is
assumed to vary with time like the trace of the energy-momentum
tensor of cosmic matter.  In the radiation-dominated era the
models reduce to standard cosmology.  In the matter-dominated era
they are, for modern values of the cosmological parameters,
consistent with data from SNe Ia searches and with the data of
\citet{GUR1999} for angular sizes of ultra compact radio sources.
We find that the angular size-redshift tests for our models offer
a higher statistical confidence than that based on SNe Ia data. A
comparison of our results with a recent revised analysis of
angular size-redshift legacy data is made,and the implications of
our models with optimized relativistic beaming in the radio
sources is discussed. In particular we find that relativistic
beaming implies a Lorentz factor less than 6,in agreement with its
values for powerful Active Galactic Nuclei.
\end{abstract}
\keywords{Cosmology:theory,dark energy,SNe Ia,angular
size-redshift relation,critical redshift,relativistic beaming}
%\newpage
\section{INTRODUCTION}

There is now substantial observational evidence\citep{PERA2003}
that favors the existence of a smooth exotic cosmic component of
energy of negative pressure.  Going at times under the name of a
cosmological constant or quintessence or,at other times,dark
energy,which we will adopt here,its true nature remains obscure.

The unexpected faintness of high redshift type Ia supernovae (SNe
Ia)suggests that the universe is accelerating
today\citep{RIE1998,PERL1999},relentlessly driven by dark energy.
When the SNe Ia results are combined with observations of the
amplitudes of primordial fluctuations in the cosmic microwave
background radiation the overall picture seems to be one of an
accelerating flat universe.  Since the standard flat universe,
despite its well-known shortcomings,has long been favored on
aesthetic and theoretical grounds\citep{KT1990},the hope has
arisen that the injection of dark energy will cure its ills,
particularly in regards of its age of the universe problem.  Thus
a major industry of investigating the constraints imposed by
continuously updated astrophysical observations on the dark energy
in refined versions of the standard model has flourished in recent
times. The present paper is one more contribution in this
direction.

When it is assumed that the dark energy,viewed in general to be
time-dependent,does not interact with matter,the energy equations
for nonrelativistic pressureless matter and dark energy decouple
leading to conservation of matter and to the dark energy equation
$d\rho_{de}/dz = 3(1+z)^{-1}(1+p_{de}/\rho_{de})\rho_{de}$,where
$\rho_{de}(z)$ and $p_{de}(z)$ are the dark energy density and
pressure respectively and $z$ is the redshift. In this case a
solvable cosmological model is obtained if a specific variation of
$\rho_{de}$ is invoked\citep{MOZ1999,AMMR2002,ARMOH2005}or a
definite parameterization for the equation of state $w(z)\equiv
p_{de}/\rho_{de}$ is suggested
\citep{CPOL2001,BACCI2003,ALAMSA2003,DR2004,PADMAN2003,CORAS2004,ALAM2004,JOHR2004,JOHR2005}.

Alternatively if one assumes that the dark energy interacts with
matter\citep{MOZ1999},the energy equations for both are coupled
and one needs a definite variation for $\rho_{de}$,in addition to
specifying its equation of state. In this case the conservation of
the energy-momentum tensor of the field equations,which holds when
matter and the dark energy are noninteracting,is replaced by the
conservation of the sum of this tensor and an extra appended
tensorial piece representing the time-dependent dark energy. Here
we follow this line: Specifically we assume (a) $\rho_{de} \sim
T$,where $T = \rho -3p$ is the trace of the energy-momentum tensor
of cosmic matter of density $\rho$ and pressure $p$,and (b) a one
parameter form for $w(z)$.

A variation $\Lambda\sim T$ was introduced by \citet{M2001,M2003}
for the cosmological constant $\Lambda$,the motivation being to
identify the cosmological constant with a Lorentz-invariant scalar
representing a form of quintessence. This cosmology is reminiscent
of  similar earlier attempts at identifying $\Lambda$ with the
Ricci scalar\citep{ALMOT1996a,ALMOT1996b,AMMR1997}. The postulate
$\Lambda \sim T$ is interesting because it implies that the
cosmological constant vanishes in the radiation-dominated cosmic
era of flat cosmology so that the successful standard primordial
nucleosynthesis predictions are unaltered. In the matter-dominated
era the postulate reduces to $\Lambda \propto H^2$ where $H$ is
Hubble's parameter. The cosmological constant variation $\Lambda
\propto H^2$ itself was widely discussed in the literature
\citep{F1987,CLW1992,LC1994,AW1994,WE1995,AR1997,OC1998,V2001}. In
particular \citet{CLW1992} have pointed out that it follows from
dimensional arguments consistent with quantum gravity. Since such
arguments do not depend on the cosmological constant equation of
state $ p_\Lambda = -\rho_\Lambda$ it is legitimate to regard them
as equally valid for dark energy with $w(z) \neq -1 $.

Extending this postulate to a dark energy with an equation of
state of negative pressure we take $\rho_{de} = \kappa T$ where
$\kappa$ is a dimensionless constant. A consequence of this is
that the matter density parameter $\Omega_m$ is constant in the
model. We take it to be $ 1/3$. This is because a matter density
parameter around 0.30 seems to be favored by observations
indicating that the dark energy accounts for 2/3 of cosmic matter
\citep{Turner2002a,Turner2002b}. In fact \citet{Turner2002c} has
strongly argued a case for $\Omega_m = 0.33 \pm 0.035 $ from
measurements of the physical properties of clusters,CMB
anisotropies and the power spectrum of mass inhomogeneities.

For the dark energy equation of state we consider 3 models with
the one-parameter forms: (1)$w_{de}=w\equiv const$,(2)$ w_{de}=
-1+ wz$,and (3)$w_{de}= -1 + w\frac{z}{1+z}$, where $w$ is
constant.

Model (1), \emph{viz}, $w_{de} \equiv w=const <0$,is a
generalization of the cosmological constant case $w_{de} =-1$.
Strictly speaking a constant $w_{de}$ is valid for the
cosmological constant only. Yet models of cosmic evolution driven
by nonrelativistic matter and a quintessence component $X$,an
exotic fluid with an arbitrary equation of state $ p_X = w_X\rho_X
$ ($w_X \geq -1 $), have been widely studied
\citep{RAP1988,CHIB1997,TUW1997,SPER1997,FRIEM1998,CALDW1998,EFST1999,Turner2002d}.
In a number of these models (particularly those with tracking
solutions),both the dark energy density parameter
$\Omega_{de}(\equiv 8\pi G\rho_{de}/3H^2)$ and $w_{de}$ vary so
slowly with redshift \citep{ZLAT1999,STE1999,EFST1999} as to
justify the approximate use of an effective equation of state
parameter
$w_{eff}\sim\frac{\int{w_{de}(z)\Omega_{de}(z)dz}}{\int{\Omega_{de}(z)dz}}$
\citep{WAN2000,ZHU2004}. More generally,the absence of robust
fundamental physics-based dark energy models and the difficulty to
observe a time dependence of $w_{de}$ from CMBR\citep{AUS2003}or
from fits to luminosity distances \citep{PICL2003},admits the
possibility of a $w_{de}$ which is constant in some specified
range,and which arises as a model-independent approximation to the
dark energy equation of state \citep{KNST2003,CEP2004}.(The
cosmology with a dark energy $ \sim a^{-2}$ and decoupled from
ordinary matter so that $w_{de} = -1/3$ has been recently
discussed by one of us \citep{AMMR2002} and by \citet{ARMOH2005}).

On the other hand parameterizations (2) and (3) are special cases
of the two-parameter forms: $w_{de} = w_{de}(0) + wz$,and $w_{de}
= w_{de}(0)+ w\frac{z}{1+z}$ which were proposed by
\citet{HUTU2001} and \citet{WEAL2002}, and  by
\citet{LIN2003}respectively, and recently studied,together with
the case $w_{de}=w\equiv const$, by\citet{DR2004}. The form
$w_{de} = w_{de}(0)+wz$ diverges at very high redshifts whereas
this difficulty is avoided in the model $w_{de} = w_{de}(0)+
w\frac{z}{1+z}$, where $w_{de} \rightarrow w_{de}(0)+w$ as
$z\rightarrow \infty$. But, as argued by \citet{Riess2004},a safer
strategy, which we follow here,is to regard these
parameterizations as only valid for low-z ($z\ll$ decoupling
redshift $z_{dec}$) and describing the late behavior of dark
energy. This was done by \citet{DR2004} who studied these
parameterizations and found,on taking as prior $\Omega_m = 0.3$ in
a flat universe,that a constant $w_{de} \equiv w_{de}(0) = -1$ is
preferred by the fit to the gold data for type Ia supernovae
\citep{Riess2004,TON2003,BARR2004}.  Also it is already known that
the cosmological constant scenario remains consistent with tight
constraints from new cosmic microwave background and galaxy
clustering data\citep{ALESS2004}. Quite generally observations
seem to require dark energy with present values $w_{de} \sim -1 $
and $ \Omega_{de} \sim 0.7$ \citep{PERA2003}. With this in mind,
and noting that recent SNeIa observations from HST do not indicate
a rapid variation of $w_{de}(z)$ away from its cosmological
constant value,we pursue,for simplicity,the following approach: we
consider the  preceding three $w_{de}(z)$ parameterizations and
set in them,$\tilde{a}b\ initio$, $w_{de}(0)= -1$. Then we
investigate the constraints on the dark energy equation of state
from recent supernova data and observations of the angular sizes
of ultra compact radio sources.

In section 2 we present the basic equations of the models. In
sections 3 and 4 we examine the constraints on them from supernova
and angular sizes data respectively. In discussing angular sizes
we compare our results with those from a recent work by
\citet{JJ2006},and also consider the implications of our angular
size-redshift relations in the presence of relativistic beaming of
the radio sources. Section 5 winds up the paper with a discussion
of the results and some concluding remarks.

\section{THE MODEL}
We consider a spatially flat FRW universe ($a$ is the RW scale
factor)
\begin{equation}\label{e1}
ds^2 = dt^2 - a^2\left(dr^2 + r^2d\theta^2 + r^2\sin^2\theta
d\phi^2\right)
\end{equation}
with cold matter of zero pressure and energy density $\rho_m$ and
dark energy of density $\rho_{de}$ and pressure $p_{de} =
w_{de}\rho_{de}$. Denoting the scale factor today by
$a_0$,(subscript $"0"$ denotes present-day quantities),and
defining $a/a_0 = (1+z)^{-1}$,where $z$ is the red-shift,
Einstein's gravitational field equations can,in this case,be
written as ($\alpha\equiv3/8\pi G$)
\begin{eqnarray}\label{e2}
\alpha^{-1}\left(\rho_m + \rho_{de} \right)&  = &H^2 \equiv H_0^2E^2(z),\\
\frac{3\alpha^{-1}w_{de}\rho_m}{(1+3w_{de})} &=& H^2
-\frac{2qH^2}{(1+3w_{de})},
\end{eqnarray}
where $q=-\frac{a\ddot{a}}{\dot{a}^2}=
\frac{\ddot{z}(1+z)}{\dot{z}^2} -2$ is the deceleration parameter
and $H=\dot{a}/a = -\frac{\dot{z}}{(1+z)}$ Hubble's constant (an
over-dot denotes time differentiation),with $H_0 \equiv100h \quad
kms^{-1}Mpc^{-1} = 2.16h\times 10^{-42} GeV$ being its present-day
value ($h$ is the normalized Hubble constant). Defining the
density parameters
\begin{equation}\label{e4}
\Omega_m\equiv\frac{\alpha^{-1}\rho_m}{H^2}, \qquad
\Omega_{de}\equiv\frac{\alpha^{-1}\rho_{de}}{H^2},
\end{equation}
we deduce from equation(\ref{e2}) that $\Omega_m + \Omega_{de}
=1$,valid at all times including $t =t_0$.

Combining equations (\ref{e2})-(\ref{e4})we obtain
\begin{equation}\label{e5}
q \equiv -1+\frac{(1+z)}{2H^2}\frac{dH^2}{dz} = \frac{1}{2} +
\frac{3w_{de}}{2}\left(1-\Omega_m\right) = \frac{1}{2} +
\frac{3\alpha^{-1}w_{de}\rho_{de}}{2H^2}.
\end{equation}

In the Einstein-de Sitter(EdeS)standard model  $\Omega_m = 1$ or
$\rho_{de} = 0$ so that $ q = 1/2$. For the cosmological constant
$\Lambda$ case $ w_{de} = -1$ so that $q=\frac{3}{2}\Omega_m -1$
which admits an accelerating universe scenario provided $ \Omega_m
< 2/3$.

In this paper we assume that $\rho_{de} = \kappa T$ where $T =
\rho -3p$ is the trace of the matter energy-momentum tensor and
$\kappa$ a dimensionless constant \citep{M2001,M2003}. Then in the
matter-dominated epoch of flat cosmology we have from equations
(\ref{e2}) and (\ref{e4}) that $\rho_{de} = \kappa \rho_{m} =
\frac{\alpha\kappa}{1+\kappa}H^2$ and $\Omega_m
=\frac{1}{1+\kappa} \equiv const$ \citep{M2001,M2003}. We further
set,as was done by \citet{M2001,M2003} and argued in the
introduction, $\Omega_m = 1/3 $. We then obtain from
equation(\ref{e5}),
\begin{equation}\label{e6}
q = -1 + \frac{(1+z)}{2H^2}\frac{dH^2}{dz} = \frac{1}{2}+ w_{de}.
\end{equation}
The rest of the paper investigates the consequences of this model
for $q$ using the different dark energy parameterizations
discussed in the introduction.
\subsection{Parameterizations of $w_{de}$}
\subsubsection{Model 1: $w_{de} \equiv w = const \leq 0$.}

Inserting $w_{de} \equiv w = const \leq 0$ in
equation(\ref{e6})yields
\begin{equation}\label{e7}
E^2(z) = (1+z)^{(3+2w)}.
\end{equation}
\subsubsection{Model 2: $w_{de} = -1 + wz,\quad w \equiv constant > 0$}

 Here equation(\ref{e6})shows that $ q
> 0, (q < 0),$ for $ z > \frac{1}{2w}, (z < \frac{1}{2w})$,implying
a cosmic deceleration - acceleration  transition at redshift $z_T
= \frac{1}{2w}$. In this case the solution of equation
(\ref{e6})for $H^2$ is ($z\ll{z_{dec}}$):
\begin{equation}\label{e8}
E^2(z) = (1+z)^{1-2w}\exp{(2wz)}.
\end{equation}
\subsubsection{Model 3: $w_{de} = -1 + w \frac{z}{(1+z)},\quad w \equiv
const > 0 $}

The deceleration-acceleration cosmic transition occurs in this
model at $z_T = \frac{1}{(2w-1)}$ so that we must have $w >
\frac{1}{2}$. In this case the solution of equation(\ref{e6}) for
$H^2$ is($z\ll{z_{dec}}$):
\begin{equation}\label{e9}
E^2(z) = (1+z)^{(1+2w)}\exp{[-\frac{2wz}{(1+z)}]}.
\end{equation}
\section{TYPE Ia SUPERNOVAE}
\subsection{The Distance Modulus}

For a flat universe the luminosity distance in units of
Megaparsecs may be defined by
\begin{equation}\label{e10}
d_L = cH_0^{-1}(1+z)\int_{0}^{z}\frac{dz}{E(z)} \equiv
c(1+z)^2d(0,z),
\end{equation}
where
\begin{equation}\label{e11}
d(z_1,z_2) = H_0^{-1}(1+z_2)^{-1}\int_{z_1}^{z_2}\frac{dz}{E(z)} .
\end{equation}
In terms of $d_L$ the predicted distance modulus is
\begin{equation}\label{e12}
\mu_p = 5\log{d_L} + 25 .
\end{equation}
We next obtain expressions for $d_L$ and $\mu_p$ in our models. In
calculating $\mu_p$ we use the widely accepted value for the
Hubble constant $H_{0} = 72kms^{-1}Mpc^{-1}$
\citep{FREE2001,FREE2003}.
\subsubsection{Model 1}

From equations(\ref{e7}) and (\ref{e10})
\begin{eqnarray}\label{e13}
d_L& \equiv& c(1+z)^2d(0,z)= \frac{2cH_0^{-1}(1+z)}{1+2w}\left[1 -
(1+z)^{-w-\frac{1}{2}}\right], \qquad w\neq -1/2,\nonumber\\
d_L&\equiv& c(1+z)^2d(0,z) = cH_0^{-1}(1+z)ln(1+z), \qquad w=-1/2.
\end{eqnarray}
Hence by equation(\ref{e12}),
\begin{eqnarray}\label{e14}
\mu_{p} &=& 43.10 + 5\log \left\{\frac{2(1+z)}{1+2w}\left[1 -
(1+z)^{-w -\frac{1}{2}}\right]\right\}, \qquad w \neq
-1/2,\nonumber\\
 \mu_{p} &=& 43.10 + 5\log\left[(1+z)\ln(1+z)\right], \qquad w =
 -1/2.
\end{eqnarray}
\subsubsection{Model 2}

From equations(\ref{e8})and(\ref{e10},
\begin{equation}\label{e15}
d_L \equiv c(1+z)^2d(0,z) =
cH_0^{-1}w^{-w-\frac{1}{2}}\exp(w)(1+z)\left[\gamma\left(w+\frac{1}{2},w(1+z)\right)-\gamma\left(w+\frac{1}{2},w\right)\right]
\end{equation}
where
\begin{equation}\label{e16}
\gamma(u,\alpha) = \int^{\alpha}_{0}t^{u-1}\exp(-t)dt ,\qquad
Reu>0,\quad |arg \alpha|<\pi .
\end{equation}
is the incomplete gamma function \citep{ABRAM1964}.  Then
\begin{equation}\label{e17}
 d_L = cH_0^{-1}\exp(w)(1+z)\int_{1}^{1+z}u^{w-\frac{1}{2}}\exp(-wu)du
\end{equation}
so that
\begin{equation}\label{e18}
\mu_p(z) = 43.10 +
5\log\left(\exp(w)(1+z)\int_{1}^{1+z}u^{w-\frac{1}{2}}\exp(-wu)du\right).
\end{equation}
\subsubsection{Model 3}

Equations(\ref{e9}) and(\ref{e10}) give $(w>1/2)$
\begin{eqnarray}\label{e19}
d_L \equiv c(1+z)^2d(0,z)
&=&cH_0^{-1}w^{\frac{1}{2}-w}\exp(w)(1+z)\left[\gamma\left(w-\frac{1}{2},w\right)-\gamma\left(w-\frac{1}{2},\frac{w}{1+z}\right)\right]\nonumber\\
&=&cH_0^{-1}\exp(w)(1+z)\int_{\frac{1}{1+z}}^{1}u^{w-\frac{3}{2}}\exp(-wu)du.
\end{eqnarray}
Hence
\begin{equation}\label{e20}
\mu_p(z)= 43.10 +
5\log\left(\exp(w)(1+z)\int_{\frac{1}{1+z}}^{1}u^{w-\frac{3}{2}}\exp(-wu)du\right).
\end{equation}
\subsection{Supernova model predictions and observations}
\subsubsection{Supernova observations}

Several astrophysics groups\citep{TON2003,BARR2004,Riess2004}have
recently updated the original supernova data of \citet{RIE1998}
and \citet{PERL1999} that provided the first glimpse into an
apparently  accelerating universe. In particular
\citet{BARR2004}have published photometric and spectroscopic
observations of 23 supernovae in the redshift range $0.3396 \leq z
\leq 1.031 $. Confronting our predictions for $\mu_p$ with their
data as analyzed by the BATM (Bayesian Adapted Template Match)
method \citep{TON2003},and calculated using $H_0 =
72kms^{-1}Mpc^{-1}$, we have minimized with respect to the
parameter $w$ the $\chi^2$ statistic:
\begin{equation}\label{e21}
\chi^2(w) =
\sum_{i=1}^{23}\frac{[\mu_{p,i}(z_i;w)-\mu_{observed,i}]^2}{\sigma_i^2},
\end{equation}
where the summation is over all 23 data points in Table 11 of
\citet{BARR2004} and $\sigma_i$ the corresponding uncertainties in
the observed distance moduli. We discuss the application of this
procedure to our 3 models.
\subsubsection{Model 1}

Using the first of equations (\ref{e14})we calculated with the aid
of equation (\ref{e21}) $\chi^2(w)$. For $w\neq -1/2$ Figure 1
shows that the resulting curve has a minimum $\chi^2_{min} = 15.4$
at $ w = -0.7$, with upper limits $w = -0.41$ and $w = - 0.28 $ at
the 68$\%$ and 95$\%$ confidence levels (c.l.) respectively (with
22 degrees of freedom- d.o.f.). For $w=-1/2$ corresponding to
$q=0$ (coasting universe) we obtain,using the second equation
in(\ref{e14}), $ \chi^2_{min}= 19.67$.

To discuss the implications of the value  $w = -0.7$ for the age
of the universe in this model we first note the following. For a
flat universe with a Hubble constant $H_0 = 72kms^{-1}Mpc^{-1}$
and contributions to the mass-energy density today of 1/3 and 2/3
of its total value from non-relativistic matter and dark energy
respectively,it is observed that the age of the universe is  $13
 Gyr$ with uncertainty of about $ \pm 1.5 Gyr$
\citep{FREE2003}.  A consistent age $t_0 = 14\pm0.5 Gyr$ is also
determined from CMB anisotropy,independently of $H_0$
\citep{KNOX2001}. Moreover,computer simulations of
Globular-cluster stars evolution produce ages of $12.5\pm1.5 Gyr$
\citep{KRCH2002}. These estimates agree with values of $t_0$
obtained by a variety of other methods, e.g. from rates of cooling
of old white dwarf stars or from radioactive chronology
\citep{OSW1996}. Finally,assuming $w_{de} = -1$ \citet{TON2003}
deduce the constraint $H_0t_0 = 0.96\pm0.04$,in agreement with the
product \citep{FREE2003}
\begin{equation}\label{e22}
\left(H_0 = 72\pm 8 kms^{-1}Mpc^{-1}\right)\times \left(t_0 = 13
\pm 1.5 Gyr \right) = 0.96\pm 0.16 .
\end{equation}
The observed ages of the universe are therefore consistent with a
consensus age of about $13\pm 1.5 Gyr$ \citep{FREE2003}.

In the present model we have,from equation(\ref{e7}),
\begin{equation}\label{e23}
H_0t_0 = \int^{\infty}_{0}\frac{dz}{(1+z)E(z)} = \frac{2}{3+2w} .
\end{equation}
Then $w= -0.7$ gives $H_0t_0 = 1.25 $.  At the 68$\%$ c.l. $w =
-0.41$ corresponding to $H_0t_0 = 0.91$, an estimate accommodated
by equation(\ref{e22}). The coasting cosmology ($w=-1/2$)
corresponds,as is well known,to $H_0t_0 =1$.
\subsubsection{Model 2}

In this model we have evaluated equation(\ref{e21})in the range $0
\leq w \leq 1 $ using equation(\ref{e18}) for the calculated
distance modulus and plotted the results in Figure 2. We note that
$\chi^2$ decreases monotonically as $w$ increases from $0$ but
reaches a minimum $\chi^2_{min} = 16.5$ at $w = 1.1$,
corresponding to the transition redshift $z_T = \frac{1}{2w} =
0.45$.  The value $w = 2.55$ corresponds to the 68$\%$ c.l. limit.
\subsubsection{Model 3}

Here we used equation(\ref{e20})in equation(\ref{e21})and plotted
$\chi^2$ versus $w: 0.55\leq w \leq 2$ in Figure 3. The curve has
a minimum $\chi^2_{min} = 16.35$ at $w = 1.7$,corresponding to the
transition redshift $z_T = \frac{1}{2w-1} = 0.42$.

\section{ANGULAR SIZE-REDSHIFT RELATION}
\subsection{General formulae}

The angular size distance of a light source is
\begin{equation}\label{e24}
d_A(z) \equiv d(0,z) = H_0^{-1}(1+z)^{-1}\int_0^z\frac{dz}{E(z)}
\equiv (1+z)^{-1}d(z) ,
\end{equation}
where $d(z)$ is the proper distance of the source.  In a flat
universe $d(z)=a_0r(z)$, where $r(z)$ is the source's radial
coordinate.

The angular size-redshift relation $\theta = \ell/d_A(z)$ where
$\theta $ is the source's angular size,and $\ell$ its intrinsic
length,measured in \emph{p}arse\emph{c}s (1\emph{pc} =
$1.542\times 10^{32} GeV^{-1}$)and assumed to be
redshift-independent,is one of observational cosmology's important
tests of cosmological models. Like other classical kinematic tests
it does not,generally,distinguish between cosmological models at
low redshifts $z\ll 1$ where the models are expected to converge.
In fact for models with constant $q$,one has,for $z\ll 1$,
\begin{equation}\label{e25}
\theta \equiv \frac{\ell}{d_A(z)} = \frac{\ell H_0}{z}\left[1 +
\frac{1}{2}(3+q)z + ...\right],
\end{equation}
which is formally the FRW result for small redshifts
\citep{SAN1988}. But for $z\geq 1$ there is less confidence in the
measurements because of possible influences of poorly understood
galactic evolutionary effects. However \citet{KELL1993} has argued
that ultra-compact radio sources with angular sizes (measured
using VLBI: Very Long Baseline Interferometry) in the
milliarcsecond (\emph{mas}= $10^{-3}\times 1^{\acute{''}} = 4.8481
\times 10^{-9}$ \emph{radians}) range (typically less than a
hundred parsecs in extent) are deeply embedded in active galactic
nuclei(AGN)\citep{BMP1997,JHK1998} and thus sheltered from
extra-galactic evolutionary effects. Objects of this type have a
fleeting existence ($\sim 100 \emph{years}$),so it is reasonable
to assume that characteristic parameters of their population (e.g.
linear sizes)do not change on a cosmological time scale.

\citet{KELL1993} showed that the angular size-redshift test for
ultra-compact sources favors the Einstein-deSitter $\Omega_m = 1$
canonical model. But subsequently \citet{JD1996,JD1997}
demonstrated that the data is compatible with low-density
constant-$\Lambda$ models,indicating that the best choice of
cosmological parameters for spatially flat universes was $\Omega_m
= 0.2$ and $\Omega_\Lambda = 0.8$.  In their latter work
\citet{JD1997},utilized  a data set of 337 ultra-compact sources
selected by \citet{GUR1994} from a 2.29 GHz survey by
\citet{PRES1985}comprising 917 sources with a correlated flux
limit of approximately 0.1 Jy (1 Jy $\equiv$ Jansky $=
10^{-26}Wm^{-2}Hz^{-1}$). From their study of this compilation
they conclude that the canonical model is ruled out by the
observed angular diameter-redshift relation. Later on
\citet{JACK2004} refined the analysis of Gurvits original data set
\citep{GUR1994} and found for flat universes that $\Omega_m = 0.24
+ 0.09/-0.07$. Building on \citet{GUR1994} earlier work
\citet{GUR1999}compiled a new data set of 330 compact radio
sources which has,subsequently,been used by several authors in
order to constrain the parameters of different quintessence
cosmological models \citep{V2001, LALCAN2002, ALCAN2002, ZHU2002,
CHEN2003, JDAL2003}.

We next give general formulae for the $\theta -z$ relations in the
present models. We write $\theta = \frac{D}{d_AH_0}$ where $ D =
6.87\times 10^{-2}\ell h$ is the source's characteristic angular
scale (in mas). These expressions are (in mas):

Model 1:
\begin{equation}\label{e26}
\theta = \frac{D(w+\frac{1}{2})(1+z)}{1-(1+z)^{-w-\frac{1}{2}}},
\qquad w\neq -1/2.
\end{equation}
\begin{equation}\label{e27}
\theta  =  \frac{D(1+z)}{\ln(1+z)}, \qquad w = -1/2.
\end{equation}

Model 2:
\begin{eqnarray}\label{e28}
\theta &=& \frac{Dw^{w+\frac{1}{2}}\exp(-w)(1+z)}{\gamma
\left(w+\frac{1}{2},w(1+z)\right) - \gamma \left(w+\frac{1}{2},w
\right)}\nonumber\\
&=&
\frac{D\exp(-w)(1+z)}{\int_{1}^{1+z}u^{w-\frac{1}{2}}\exp(-wu)du}.
\end{eqnarray}

Model 3:
\begin{eqnarray}\label{e29}
\theta &=& \frac{Dw^{w-\frac{1}{2}}\exp(-w)(1+z)}{\gamma
\left(w-\frac{1}{2},w \right)- \gamma
\left(w-\frac{1}{2},\frac{w}{1+z} \right)}\nonumber\\
&=&
\frac{D\exp(-w)(1+z)}{\int_{\frac{1}{1+z}}^{1}u^{w-\frac{3}{2}}\exp(-wu)du}.
\end{eqnarray}
Using  $\partial\gamma(u,\alpha)/\partial\alpha =
\alpha^{u-1}\exp(-\alpha)$  we find the small$-z$ expansions of
these equations:

Model 1:
\begin{equation}\label{e30}
\theta = \frac{D}{z}\left(1 +\frac{1}{2}(3+q)z+...\right),\qquad
w\neq -1/2,
\end{equation}
\begin{equation}\label{e31}
\theta = \frac{D}{z}\left(1+\frac{3}{2}z+...\right), \qquad w=
-1/2,
\end{equation}
 where from equation(\ref{e6}) $q= \frac{1}{2}+w$. Thus both these
 equations
formally coincide with equation(\ref{e25}).

Models 2 and 3:
\begin{equation}\label{e32}
\theta = \frac{D}{z}\left[1+\frac{5}{4}z +
(9+8w)\frac{z^2}{48}+...\right],
\end{equation}
where $q = -\frac{1}{2} +wz$ in model 2 and $q = -\frac{1}{2} +
w\frac{z}{1+z}$ for model 3. In both cases the small-$z$ expansion
of $\theta$ agrees with equation(\ref{e25}) on retaining only the
$w$-independent part of $q$.

\subsection{Critical redshift}

The existence of a critical redshift $z_m$ corresponding to a
minimum angular size can be qualitatively understood as follows.
The reason for it is not just because,in the context of cosmic
expansion,was the light received today from a source emitted when
the source was closer,but also,more importantly,because at a
larger $z$ the source of a standard linear size occupies a larger
fraction of a large circle. From equations (\ref{e26})-(\ref{e29})
the redshift $ z_m$ satisfies the equations:

  Model 1:
\begin{equation}\label{e33}
z_m = \left(w+ \frac{3}{2}\right)^{\frac{1}{w+\frac{1}{2}}} -
1,\qquad w\neq -1/2,
\end{equation}
\begin{equation}\label{e34}
 z_m = e -1 \simeq 1.72, \qquad w = -1/2.
 \end{equation}

  Model 2:
  \begin{equation}\label{e35}
  \left(1+z_m\right)^{w+\frac{1}{2}} =
  \exp\left[w\left(1+z_m\right)\right]\int_1^{1+z_m}u^{w-\frac{1}{2}}\exp(-wu)du.
  \end{equation}

  Model 3:
  \begin{equation}\label{e36}
  (1+z_m)^{-w+\frac{1}{2}}=
  \exp\left[\frac{w}{1+z_m}\right]\int_{\frac{1}{1+z_m}}^1u^{w-\frac{3}{2}}\exp(-wu)du.
  \end{equation}

Several authors \citep{KRAU1993, LIMA2000a, LIMA2000b, JDAL2003}
studied critical redshifts in different models to find out how
sensitive $z_m$ is to variation of parameters like $w$. We address
this question in \S4.3.2.
\subsection{Constraints from angular size measurements}
\subsubsection{$\chi^2$ analysis}

Our object now is to investigate constraints on the parameters $w$
and $D$ using the data compilation of \citet{GUR1999} for the
angular size measurements of milliarcsecond radio sources,observed
at frequency $\nu = 5 GHz$, in the redshift range $0.011\leq z
\leq 4.72.$ The number of sources, originally 330, was reduced to
those with a spectral index $\alpha$ in the range $-0.38\leq
\alpha \leq 0.18$ and total luminosity $ L \geq 10^{26} W/Hz$ so
as to minimize any possible dependence of angular size on $\alpha$
and also restrict the intrinsic size of the sources.  These
criteria were met by 145 sources which were then grouped into 12
bins of 12-13 sources per bin. This binned data was used in Figure
10 of the paper by \citet{GUR1999}.

We attempted the determination of the best values of the models
parameters $ D$ and $w$ through a simultaneous minimization with
respect to $D$ and $w$ of the $\chi^2$ statistic
\begin{equation}\label{e37}
\chi^2(w,D)=
\sum_{i=1}^{12}\frac{\left[\theta_{p,i}(z_i;w;D)-\theta_{measured}(z_i)\right]^2}{\sigma_i^2},
\end{equation}
where $\theta_{p,i}$ denote the predicted angular sizes given by
equations (\ref{e26})-(\ref{e29}),$\theta_{measured}(z_i)$ the
corresponding observed values of \citet{GUR1999},and $\sigma_i$
the observation's error of the sample's $ith$ bin. We have chosen
the range of $D$ to span the interval $[0.10,2.20]  \emph{mas}$
and $w$ in intervals appropriate to each model. We adopt a
minimization procedure,for 10 d.o.f,that produces minimum $\chi^2$
values for $w$ values corresponding to different values of $D$ in
its range of variation. The least value among these $\chi^2$
minima,denoted $\chi^2_{min}$,corresponds to a minimization of
$\chi^2$ with respect to both $D$ and $w$. The results are
displayed in Tables 1, 2, and 3, and in Figure 5. The values of
$w$ and $D$ corresponding to $\chi^2_{min}$ are as follows.

\textbf{Model} \textbf{1}

  \emph{Case} $w\neq -1/2$:

 Using equation (\ref{e26}) in equation(\ref{e37})with $w$ in the
 range $[-1,0]$ we obtain (Table 1) $\chi_{min}^2=4.72$ corresponding to $w = -0.45$ and
  $D= 1.25 \emph{mas}$. Figure 4 shows, for $-1\leq w \leq 0$, the $68\%$ and $95\%$ confidence contours in the $w-D$
plane.

 \textbf{Model} \textbf{1}

 \emph{Case} $w = -1/2$ (\emph{Coasting cosmology}):

 From equations (\ref{e27}) and (\ref{e37}) we deduce (see Figure 5 ) that
 $\chi^2_{min} = 4.75  $ at $ D = 1.3^{+0.27}_{-0.29}$ \emph{mas} and $D=
 1.3^{+0.40}_{-0.43}$ \emph{mas}, where the errors produce the $68\%$ and
 $95\%$ confidence limits respectively.

 \textbf{Model 2}

 Using equations (\ref{e28}) and (\ref{e37}) and taking $w$ in
the range [0,3] we note,in this case,that $\chi^2$ decreases
monotonically from $\chi^2 = 104.29$ at $D = 0.10 \emph{mas}$ and
$w = 3$ to the minimum value $\chi^2_{min} = 4.47$ at $D = 1.45
\emph{mas}$ and $w= 0.50$ ( Table 2). The value $w = 0.50 $
corresponds to $z_T = 1.00$. The $68\%$ and $95\%$ confidence
contours in the $w-D$ plane,for $0 \leq w \leq 3$, are displayed
in Figure 6 .

\textbf{Model 3}

In this model equations (\ref{e29}) and (\ref{e37}) reveal,for $w$
in the range $[0.55,3]$, that $\chi^2$ descends from $\chi^2 =
108.87$ at $D= 0.10 \emph{mas}$ and $w = 3$ to a minimum
$\chi^2_{min} = 4.58$ at $D = 1.40 \emph{mas} $ and $w = 1.15$ (
Table 3).  This value of $w$ corresponds to $z_T = 0.77$. For $
0.55 \leq w \leq 3$, the $68\%$ and $95\%$ confidence contours are
shown in Figure 7.
\subsubsection{Calculation of the critical redshifts}

Equations (\ref{e33})-(\ref{e36}) can be solved for $z_m$
corresponding to specific values of $w$. The results are plotted
in Figures 8-10 for models 1-3 respectively. The redshifts $z_m$
for the best-fit $w$ values are shown on the diagrams.  We note
that in both models 1 and 3 the best-fit $w$ value corresponds to
$z_m = 1.65$.  This reiterates the point made by \citet{JDAL2003}
that the minimum redshift test "cannot by itself differentiate
between different cosmological models" because different scenarios
might correspond to the same $z_m$. The first and third models in
this paper are generically quite distinct, yet they have the same
$z_m$.  In fact for all three models the values of $z_m$ are quite
close.  In particular in model 2 the best-fit $w$ gives $z_m =
1.71$ which almost coincides with $z_m = 1.72$ for coasting
cosmology. Expectedly for $w =0$ in model 1 we recover (Figure 8)
the standard model result $z_m = 5/4 $.

\subsubsection{Comparison with \citet{GUR1999}data}

The predictions of the $\theta -z$ relations (\ref{e26})-
(\ref{e29}) for $D$ and $w$ corresponding to $\chi^2_{min}$ in
each model are plotted,alongside the data  of \citet{GUR1999},in
Figure 11. ( Table 4 gives $\log z$ and the corresponding
$\log\theta$ values for \citet{GUR1999} data points and the end
points $ \log\theta_\pm$ of their error bars). The inset magnifies
the neighborhood of the minima of the curves. The curves,drawn for
the best-fit values of $w$ and $D$,cluster within a narrow spread
of $\log\theta$ and appear to be reasonably consistent with
\citet{GUR1999} data. None of the four models appears to be
particularly favored over the others by this data.

Lastly in the work of \citet{GUR1999} a multi-parameter regression
analysis of the data with $\ell h$ a free parameter yields $\ell h
= 23.8\pm 17.0 \emph{pc}$. This corresponds to $D=1.64\pm 1.17
mas$, a range within which fall all the $D$ values in Table 5.

\subsection{Credibility of ultra compact radio sources as
standard measuring rods}

In his 1993 analysis of compact radio sources
\citet{KELL1993}found the angular sizes to be essentially
independent of redhift in the interval $0.5 <z<3$. In the latter
work of \citet{GUR1999}this feature appeared to persist for median
 angular sizes of sources with $z>0.5$. Although according to
\citet{KELL1993},and as mentioned earlier on,it is reasonable to
assume that milliarcsecond ultra-compact radio sources are not
affected by evolutionary effects,it is still important to treat
\citet{GUR1999} data ,as emphasized by them, with caution. A
widely-accepted model of an ultra compact radio source is one in
which a central low luminosity object is straddled by a pair of
radio-bright lobes emitting synchrotron radiation. The lobes are
sustained by two hot gas jets\citep{PEAC1999,BMP1997,JHK1998}. In
such a picture differences in the spectral index between the
core,the central engine, and its jet components may introduce a
"linear size-redshift " dependence even in the absence of
evolution. \citet{FRE1997}have however argued that such a
dependence is likely to be weak. Nevertheless more data at various
frequencies going beyond limited source samples can,in
particular,help verify the importance of such a dependence and
also,more generally,enhance further the credibility of
milliarcsecond radio-sources as standard measuring rods. With this
goal in mind \citet{JJ2006}have recently studied the implications
of an updated sample of sources in the \citet{PRES1985}catalogue.
We shall shortly discuss their work and compare their results to
ours. Another aspect of interest is relativistic
beaming\citep{PEAC1999,BMP1997,JHK1998},the process by which the
D\"{o}ppler effect modifies the appearance of a radio source with
lobes. For a lobe whose jet axis is oriented close to the line of
sight from the source to the earth relativistic beaming generally
increases the apparent radio power of the source and decreases its
angular size. The implications of this orientation bias on the FRW
$\theta -z$ relation\citep{SAN1988}have been studied by
\citet{DAB1995}with the aim of evaluating the statistical
confidence in the observational data,particularly in relation to
estimates of the value of $q_0$. A similar study was also
undertaken by \citet{STEP1995}. We shall consider a simple
variation of the \citet{DAB1995}approach in order to assess the
impact of this orientation bias on our calculations.

\subsubsection{Comparison with the work of \citet{JJ2006}}

\citet{JJ2006}reexamined the \citet{PRES1985} catalogue of ultra
compact radio sources,updating it with respect to both red shift
and radio information and replacing the original choice of 337
sources by \citet{GUR1994}with a sample of 613 objects in the red
shift range $z = 0.0035$ to $z = 3.787$.

Over its red shift range of $0.003\leq z\leq 3.8$ \citet{GUR1994}
337 objects sample contains sources of luminosity that varies over
three orders of magnitude,from 0.01 to 1 Jy (see Fig 3 in
\citep{GUR1994}). Moreover the sources in the high red shift range
$z> 0.5$ exhibit a smaller dispersion in luminosity as compared to
sources with $z<0.5$. This suggests that the higher red shift
objects have similar linear sizes. This is why \citet{GUR1994},
and also \citet{JJ2006},ignore objects with $z < 0.5$.

This mismatch between the luminosity of higher and lower z sources
is also apparent in the 330 sources sample of \citet{GUR1999}(see
Fig 3 in \citet{GUR1999}). This is why that we also limit
ourselves in this paper to \citet{GUR1999} data with $ z > 0.5$
(see Fig 10 in \citet{GUR1999} and Table 4 in the present paper).

Of the 613 sources in the sample of \citet{JJ2006} 468 have $z >
0.5 $. These were placed in 6 bins with 78 objects in each bin.
But \citet{JJ2006} show that the cosmological parameters which
best fit the $\theta-z $ data in this case,including a
best-fitting value $\ell = 7.75h^{-1}kpc$ for the source's
intrinsic length,are not very sensitive to the choice of binning.
Rather than relying too heavily on a particular binning
\citet{JJ2006} produce data points corresponding to bin sizes of
76, 77 and 78 objects and take for their $(z, \theta)$ points
values that are a composite of these three cases. Their resulting
data points are: $(z, \theta (mas)) \equiv (0.6153,1.4624),
(0.8580, 1.2801),(1.1527, 1.1599),(1.4200, 1.1448),$
$(1.8288,1.1760),$ and $ (2.5923,1.2374)$,with standard deviation
$\sigma = 0.00603$. Using this data,and marginalizing over the
sources intrinsic length,\citet{JJ2006} obtain the best-fit
parameter values $\Omega_m = 0.25 + 0.04/-0.03, \Omega_\Lambda =
0.97 + 0.09/- 0.13$ and the curvature parameter $k = 0.22 +
0.07/-0.10$,where the error bars are $68\%$ confidence limits.
These results do not agree with our models. In fact the
$\log\theta-\log z$ curve for\citet{JJ2006} results would lie well
below the base of Figure 11. We have confirmed this disagreement
by doing a $\chi^2$ analysis of the predictions of our models
against \citet{JJ2006} angular size results. We will comment on
this matter in the summary and conclusions section.

\subsubsection{Effect of orientation of lobe jet axis on $q_0$:Method of \citet{DAB1995}}

\citet{DAB1995} gave qualitative and quantitative discussions of
the effect of relativistic beaming on the FRW $\theta -z$
relation. They consider a simulation sample of compact radio
sources,with each source composed of two identical but oppositely
directed jets having an angle $\phi$ to the line of sight,where
$0<\phi\leq \pi$. Each source emits an isotropic power-law
spectrum with flux density
\begin{equation}\label{e38}
S =
\frac{L}{d_L^2}\left[\frac{1}{\gamma(1-\beta\cos\phi)}\right]^{3+\alpha},
\end{equation}
where $L$ is the source's luminosity, $d_L$ is the luminosity
distance, $\alpha$ is the spectral index, $\beta$ is the jet
speed/c,and $\gamma = (1-\beta^2)^{-1/2}$ is the Lorentz factor of
the jets. For a quasi-continuous jet formed out of finite-lifetime
blobs the appropriate power index above is reduced to $(2+\alpha)$
\citep{PEAC1999}.

The inclusion of the source's lobe orientation in the FRW $\theta
-z$ relation
\begin{equation}\label{e39}
\theta = \frac{\ell H_0q_0^2(1+z)^2}{q_0z +
(q_0-1)\left((2q_0z+1)^{1/2}-1\right)}
\end{equation}
replaces the source's intrinsic length $\ell$ by
$\ell_{\perp}=\ell\sin\phi$ where $\ell_{\perp}$ is the projection
of $\ell$ in the plane of the sky ($\ell_{\perp}(\phi
=\frac{\pi}{2})=\ell,$ for a source whose jet axis lies along the
plane of the sky)

It follows from equation (\ref{e38})that the maximum allowed angle
in the sample is given by \citep{DAB1995}:
\begin{equation}\label{e40}
\phi_{max}(z) =
\arccos\left(\frac{1}{\beta}-\frac{1}{\gamma\beta}\left[\frac{L}{S_{lim}d_L^2}\right]^{\frac{-1}{3+\alpha}}\right),
\end{equation}
where $S_{lim}$ denotes the sources flux density limit.

 \citet{DAB1995}define the average lobe orientation angle by
\begin{equation}\label{e41}
\overline{\phi(z)}= \frac{\int_0^{\phi_{max}}\phi\sin\phi
d\phi}{\int_0^{\phi_{max}}\sin\phi d\phi}.
\end{equation}
Since $\phi$ is small one can assume $\overline{\ell_\perp(z)} =
\ell\sin(\overline{\phi(z)})$ and insert
$\overline{\ell_\perp(z)}$ into the $\theta -z$ relations. The
essence of the \citet{DAB1995} quantitative method is to
investigate the effect of this modification on the predictions of
equation (\ref{e39}).

\subsubsection{Optimized relativistic beaming: an alternative
approach to \citet{DAB1995} calculation}

The twin-jet model of \citet{DAB1995} assumes that the measured
angular size of the source corresponds to the separation of the
jets projected onto the plane of the sky. \citet{JACK2004}
considers this feature to be unrealistic because the flux of the
receding (away from us) jet is very much smaller than that of the
advancing(towards us)jet,for example by a factor of up to $10^6$
for a jet Lorentz factor $\gamma$ of 5, and may therefore be
neglected \citep{LB1985}.

The relativistic beaming hypothesis is primarily based on two
parameters: the Lorentz factor $\gamma$ (or equivalently the
parameter $\beta \equiv$ the jet velocity/c ) and the viewing
angle $\phi$. In a study of 25 quasars with $z$ between 0.1 and
2.5 (some quasars with their nonstellar radiation spectrum and
evidence of jets and extended emission features also display
relativistic beaming effects)\citet{VERC1994} find that $\gamma$
increases with $z$ if $q_0 = 0.05$ but remains constant at $\gamma
\approx 10$ if $q_0 = 0.5$. With this in mind \citet{DAB1995}
take, in their analysis of a simulation data, $\gamma = 10$,
although they point out that this choice does not qualitatively
affect their results.

However a choice of $\gamma$ linked to a positive $q_0$,as in the
1995 paper by \citet{DAB1995},is hard to justify in these days
where $q_0 < 0$ (accelerating universe) is very probable. Moreover
there are other reasons that might render an arbitrary choice of
$\gamma$ fraught with uncertainties\citep{UBCH2002}. In addition,
geometric projection effects of components of jets of galaxies or
quasars can lead to dramatically misleading values of $\gamma$.
For example the transverse motion of blobs of gas across the sky
in a well-studied jet in the galaxy $M87$ ( at about $18Mpc$ away
from us)appears to be \emph{superluminal}, corresponding to a
velocity nearly 6 times the speed of light!\citep{CHAM2002}. This
cosmic illusion of faster-than-light velocities is simply a
projection effect, produced by the blobs of gas moving at a very
small angle $\phi$ at a near-velocity-of-light speed
\citep{CHAM2002}. It is worthy of mention that Doppler boosting
was first discussed by \citet{SHK1964a,SHK1964b}to explain the
apparently one-side jet in$M87$.

In this paper we choose to relate $\gamma$ to  $\phi$ by
optimizing the relativistic beaming effect. This technique was
used by \citet{VERC1994},\citet{UBA1999},and \citet{UBCH2002}.

Define the Doppler factor which fundamentally characterizes
relativistic beaming by (see equation (\ref{e38})):
\begin{equation}\label{e42}
\delta = [\gamma(1-\beta\cos\phi)]^{-1}
\end{equation}
Optimization of relativistic beaming is obtained by requiring
$\frac{d\delta}{d\beta} =0$. This yields
\citep{VERC1994,UBA1999,UBCH2002} $\phi=\arccos\beta$, or
$\phi=\arcsin\frac{1}{\gamma}$.  Since the lobe orientation bias
replaces $\ell$ by $\ell\sin\phi$ in the $\theta -z$ relation,
$D=6.87\times10^{-2}\ell h$ in equations (\ref{e26})-(\ref{e29})is
replaced by $\frac{D}{\gamma}$. In what follows we discuss the
constraints of the observational data on $\frac{D}{\gamma}$.

Clearly application of the $\chi^2$procedure of \S (4.3.1.) to the
modified $\theta -z$ relations using the binned \citet{GUR1999}
data assigns now the best-fit values of $D$ to $\frac{D}{\gamma}$
and leaves those of $w$ (and hence of $q_0$)undisturbed. Taken
together the  three models imply (see Table 5) $1.25 \leq
\frac{D}{\gamma} \leq 1.45 $ mas, or, equivalently,
$18h^{-1}\gamma \leq \ell \leq 21h^{-1}\gamma$ pc.  Since
milliarcsecond ultra-compact radio sources are typically less than
a hundred parsec in extent the last inequality would imply that
$18h^{-1}\gamma \leq \ell < 100 pc$, or $\gamma < \frac{100}{18}h
\leq \frac{100}{18} < 6$. (If we use $h=0.72$ we get $\gamma <
4$). It is known \citep{AEL1990,PEAC1999}that the most powerful
AGN jets seem to have Lorentz factors of the order of these upper
bounds.

\section{SUMMARY AND CONCLUSIONS}

We have studied in this paper supernova and compact radio sources
angular sizes constraints on three cosmological models whose
dynamics is driven by non-relativistic matter of density parameter
$\Omega_m = \frac{1}{3}$ and a smooth time-dependent dark energy
component with density $\rho_{de} \propto T = \rho - 3p $ ($T$ is
the matter energy-momentum tensor) and equation of state $w_{de} =
\frac{p_{de}}{\rho_{de}}, w_{de}$ being either constant or
redshift-dependent, with $w_{de}(z=0) = -1 $ in the latter case.
The variation $\rho_{de} \sim T$ implies that the dark energy
vanishes in the early universe leaving the standard model's
primordial nucleosynthesis predictions intact. This is somewhat
reminiscent of the modified general relativity model of
\citet{ALMOT1996a,ALMOT1996b} which can be cast in the form of a
variable-$\Lambda$ cosmology with $\Lambda \propto R$, $R$ being
the Ricci tensor. There the radiation density $\rho_r \sim a^{-4}
\sim t^{-2}$ in the early universe, as in standard flat cosmology.
However the Friedmann equation is modified so that the standard
cosmic expansion rate is altered by the factor
$\frac{1}{3}(1+(2/\eta))$ where $\eta$, a constant, satisfies $
0\leq \eta \leq 1, \eta = 1$ in general relativity.  Here the
postulated dark energy ansatz does not affect the standard cosmic
expansion rate of the early universe.  It reduces in the
matter-dominated phase to $\rho_{de} \sim H^2$,a variation that
was extensively studied for the cosmological constant
\citep{OC1998}. Our main results are summarized in Table 5.

For the first model with $w_{de}$ constant ( implying by equation
(\ref{e6}) that $q \equiv q_0 = constant$) the supernova data of
\citet{BARR2004} admits an accelerating universe with $w_{de} =
-o.7 $ and $q_0 = -0.2$ ($\chi_{min}^2 = 15.4$) and a coasting
universe with $w_{de} = -0.5$($\chi_{min}^2 = 19.67$). On the
other hand \citet{GUR1999} angular sizes data allows a mildly
decelerating universe with $w_{de} =-0.45$,$q_0 = 0.05$ and
characteristic angular scale $ D = 1.25 \emph{mas}$ ($\chi_{min}^2
= 4.72$) and also  a coasting cosmology with $w_{de} = - 0.50$ and
$D= 1.30 \emph{mas}$ ($\chi_{min}^2 = 4.75$).   The value $D=1.3
\emph{mas}$ corresponding to a source's intrinsic length $\ell =
18.92h^{-1} \emph{pc}$ is very close to $D=1.28$ obtained as the
best-fit value in an earlier model by \citet{JDAL2003} where the
scale factor is essentially linear in  $t$,\emph{viz},$ a\sim
t^{1.006}$.

 The results of the constant-$w_{de}$ model (and also those of the
 other models in this paper) do not extrapolate to the radiation
 dominated universe since there,as we have pointed out in the
 introduction, $\rho_{de}$ vanishes. Nevertheless an accelerated
 expansion in model 1 may pose problems for structure condensation
 after matter dominance: structure tends not to form in the
 presence of cosmic acceleration. However  the modest
 acceleration ($q = q_{0} = - 0.2$) in model 1 corresponds to
 $w_{de} = -0.7$ with upper limits $w_{de} = -0.41$ and $w_{de} =
 - 0.28$ at the $68\%$ and $95\%$ confidence levels respectively.
 These values of $w_{de}$ correspond from equation (\ref{e6}) to $
 q = 0.09$ and $ q = 0.22$ respectively so that within the limits
 of the quoted confidence levels a decelerating universe is not
 excluded.  In fact it has been argued \citep{VISH2003} that absorption
 by intergalactic dust of light travelling over immensely long
 distances might explain the faintness of extragalactic SNe Ia
 obviating the need for a cosmic acceleration
 based explanation.  Yet we are inclined to believe that our supernova
 results for the coasting universe scenario are more robust
 because of their concordance with those of \citet{JDAL2003} which
 are obtained by a different approach. A flat cosmology with
 dark energy of constant $w_{de}$ and matter density
 parameter $\Omega_m$ held at $\Omega_m = 0.3$ was also considered
 by \citet{DR2004} who however assumed noninteracting dark energy
 and used the larger set of the supernova "gold" data. They find
 in this case the preferred (cosmological) constant value $ w_{de}
 = -1$.

Models 2 and 3,which by construction describe a present-day
accelerating universe with $q_0 = -0.5$,are consistent with the
data for universes that undergo deceleration-acceleration
transitions at redshifts in the range $ 0.42 \leq z_T \leq 1$.  In
the standard $\Lambda CDM$ cosmological model $ z_T = 0.67$ for
$\Omega_m \sim 0.30$, in contrast with the observational value
$z_T = 0.46\pm 0.13$ from the SNeIa analysis of \citet{Riess2004}.
Predictions of $z_T$ in seven popular quintessence models inspired
by supergravity or M1 string theory have recently been studied by
\citet{GARD2005} who noted that all of them can, in the low-$z$
approximation $w_{de} \approx w_0 +w_1z$,($0\leq z\leq 5$),mimic
the $\Lambda CDM$ model. Here in models 2 and 3 the SNeIa data
give $z_t = 0.45$ and $z_T = 0.42$,both values being very close to
the observational result $z_T = 0.46$ of \citet{Riess2004}. On the
other hand consistency of the models with the angular size data
\citep{GUR1999}  yield $z_T =1$ and $z_T = 0.77$, to be compared
with  $z_T = 0.67 $ from the $\Lambda CDM$. Inspection of Figure
11 and Table 5 leads to an important conclusion:The model's
angular size-redshift curves drawn in Figure 11 fit
\citet{GUR1999} data equally well. This is a reflection of the
small values and span of $\chi_{min}^2$ as seen in Table 5 for the
results : for the four models $\chi_{min}^2$ varies only from 4.47
to 4.75. By comparison in the supernova test the $\chi_{min}^2$
values are considerably larger and lie in the range 15.4 to 19.67
so that one has more faith in the agreement of the models with the
angular size data.

The predictions of the models for the critical minimum redshift in
the angular size-redshift relation give $z_m$ values in the narrow
interval [1.65,1.72] compared to the standard model result $z_m =
1.25$. The constant-$w_{de} \neq -1/2$ and the third models have
the same $z_m = 1.65$ whereas the coasting universe and model 2
have $z_m 1.72$ (the same as that of \citet{JDAL2003})and $z_m =
1.71$ respectively. Thus the minimal redshift cannot,by
itself,effectively discriminate between these models.

We have also compared the predictions of our models with those of
the recent work of \citet{JJ2006}in which they revisit the old
\citet{PRES1985}catalogue of ultra compact radio sources and
reconsider an angular size-redshift data set in the light of
modern preferences of the  cosmological parameters. We find that
their results do not agree with our models. In a sense this is not
surprising since the underlying premises of the two works are
different:\citet{JJ2006}approach is based on using one simple
potential to test the hypothesis that vacuum energy is constant.
The present models are phenomenological and based on a
time-dependent dark energy coupled to matter. Hence there is no
overlap between the two approaches\citep{JJ2007}.

Finally,we tested our angular size-redshift relations in the
presence of  relativistic beaming. Relativistic beaming in the FRW
$\theta -z$ relation was investigated by \citet{DAB1995}. Here we
used a simple variation of their method in a two-jet model of
radio sources with the advancing (towards us)jet axis close to the
line of sight. In this picture consistency of the models with
\citet{GUR1999} data is found for a beaming Lorentz factor $\gamma
< 6$. Such an upper bound is consistent with values of the Lorentz
factor for powerful AGN jets \citep{AEL1990,PEAC1999}.

To conclude, we have presented dark energy models that are in
reasonable agreement with the supernova data of
\citet{BARR2004}and in good agreement with the \citet{GUR1999}
compact radio source angular size versus redshift binned data. The
three models that we have studied are simplified versions of ones
recently considered by \citet{DR2004} in a different context. But
as remarked by these authors comparing models for the equation of
state of dark energy will remain something of a mug's game until
there exists substantially more data at higher redshifts.

We thank R.G. Vishwakarma for sending \citet{GUR1999} complete
data sets on angular sizes of compact radio sources,and
J.C.Jackson for a useful correspondence. We also acknowledge the
financial support of the Directorate of Scientific Research of the
University of Khartoum. One of us (AMMAR) thanks Professor K.R.
Sreenivasan for hospitality at the Abdus Salam International
Center for Theoretical Physics, Trieste, Italy, where part of this
work was done , and the Swedish Agency for International
Development (Sida) for financial support. AMMAR also thanks Dr.
Abu Bakr Mustafa for his hospitality at ComputerMan College for
Computer Studies, Khartoum, where also part of this work was done.
%\newpage

\clearpage
\begin{deluxetable}{ccc||cc||ccc} \tablewidth{0pt}
\tablecaption{$\chi^2 - w- D$ values for Model 1}
 %\tabletypesize{\scriptsize}
\tablehead{\colhead{D}&\colhead{$w$}&\colhead{$\chi^2$}&\colhead{}&\colhead{}&\colhead{D}&\colhead{$w$}&\colhead{$\chi^2$}}
 \startdata
 0.1 &0&106&&& 1.25&-0.45&4.72\\
 0.2 &0&85.44&&&1.30&-0.55&4.74\\
 0.3&0&66.53&&&1.35&-0.60&4.75\\
 0.4&0&50.16&&&1.40&-0.65&4.80\\
 0.5&0&36.33&&&1.45&-0.70&4.88\\
 0.6&0&25.03&&&1.50&-0.80&4.97\\
 0.7&0&16.27&&&1.55&-0.85&5.08\\
 0.8&0&10.04&&&1.60&-0.90&5.20\\
 0.9&0&6.36&&&1.65&-0.95&5.35\\
 0.95&0&5.47 &&&1.70&-1&5.51\\
 1.00&-0.05&5.17&&&1.80&-1&6.28\\
 1.05&-0.15&5.02&&&1.90&-1&7.96\\
 1.10&-0.20&4.90&&&2.00&-1&10.54\\
 1.15&-0.30&4.80&&&2.10&-1&14.02\\
 1.20&-0.40&4.76&&&2.20&-1&18.41\\
\enddata
\end{deluxetable}
%\newpage
\begin{deluxetable}{ccc||cc||ccc}
\tablewidth{0pt}  \tablecaption{$\chi^2 - w- D$ values for Model
2}
 %\tabletypesize{\scriptsize}
\tablehead{\colhead{D}&\colhead{$w$}&\colhead{$\chi^2$}&\colhead{}&\colhead{}&\colhead{D}&\colhead{$w$}&\colhead{$\chi^2$}}
\startdata
 0.1&3&104.29&&&1.25&1.00&5.00\\
 0.2&3&80.97 &&&1.30&0.90&4.78\\
 0.3&3&60.91&&& 1.35&0.70&4.61\\
 0.4&3&44.12&&& 1.40&0.60&4.50\\
 0.5&3&30.59&&&1.45&0.50&4.47\\
 0.6&3&20.33 &&&1.50&0.40&4.52\\
 0.7&3&13.32 &&&1.55&0.30&4.64\\
 0.8&3&9.58 &&&1.60&0.20&4.84\\
 0.85&2.80&8.85 &&&1.65&0.15&5.07\\
 0.9&2.50&8.24 &&&1.70&0.10&5.39\\
 0.95&2.20&7.65 &&&1.75&0&5.78\\
 1.00&1.90&7.08 &&&1.80&0&6.28\\
 1.05&1.70&6.56  &&&1.90&0&7.96\\
 1.10&1.50&6.09&&&2.00&0&10.54\\
 1.15&1.30&5.66&&&2.10&0&14.02\\
1.20&1.10&5.30&&&2.20&0&18.41
\enddata
\end{deluxetable}
%\newpage
\begin{deluxetable}{ccc||cc||ccc} \tablewidth{0pt}
\tablecaption{$\chi^2 - w- D$ values for Model 3}
 %\tabletypesize{\scriptsize}
\tablehead{\colhead{D}&\colhead{$w$}&\colhead{$\chi^2$}&\colhead{}&\colhead{}&\colhead{D}&\colhead{$w$}&\colhead{$\chi^2$}}
\startdata
 0.1&3&108.87&&&1.25&1.85&4.83\\
 0.2&3&89.00&&&1.30&1.55&4.70\\
 0.3&3&71.25&&&1.35&1.35&4.61\\
 0.4&3&55.63 &&&1.40&1.15&4.58\\
 0.5&3&42.13&&&1.45&0.95&4.61\\
 0.6&3&30.77&&&1.50&0.75&4.69\\
 0.7&3&21.53 &&&1.55&0.65&4.82\\
 0.8&3&14.42 &&&1.60&0.55&5.04\\
 0.9&3&9.43 &&&1.70&0.55&6.28\\
 1.00&3&6.57 &&&1.80&0.55&8.59\\
 1.05&3&5.94 &&&1.90&0.55&11.97\\
 1.10&2.65&5.80 &&&2.00&0.55&16.43\\
 1.15&2.35&5.28 &&&2.10&0.55&21.95\\
 1.20&2.05&5.03 &&&2.20&0.55&28.55\\
\enddata
\end{deluxetable}
%\newpage
\begin{deluxetable}{cccccc} \tablewidth{0pt}
\tablecaption{Angular size-redshift data \citep{GUR1999}}
 %\tabletypesize{\scriptsize}
\tablehead{\colhead{}&\colhead{$log z$}&\colhead{$log
\theta$}&\colhead{$\sigma$}&\colhead{$log \theta_+$}&\colhead{$log
\theta_-$}} \startdata
  1&-0.29&0.785&3.216&0.961&0.461\\
  2&-0.20&0.703&2.346&0.840&0.420\\
  3&-0.12&0.585&1.504&0.722&0.374\\
  4&-0.05&0.550&3.170&0.823&-0.427\\
  5&0.04&0.597&1.277&0.704&0.411\\
  6&0.10&0.550&2.683&0.765&-0.075\\
  7&0.15&0.562&0.634&0.621&0.482\\
  8&0.19&0.332&1.354&0.537&-0.107\\
  9&0.29&0.389&0.895&0.508&0.196\\
  10&0.34&0.690&1.078&0.768&0.573\\
  11&0.41&0.525&0.734&0.623&0.433\\
  12&0.55&0.633&0.990&0.731&0.535
\enddata
\end{deluxetable}
%\newpage
\begin{deluxetable}{c|ccccc|ccccccc|c} \tablewidth{0pt}
\tablecaption{Summary of the results}
 %\tabletypesize{\scriptsize}
\tablehead{\colhead{Model}&\colhead{}&\multicolumn{4}{c}{SNeIa}&\colhead{}&\multicolumn{5}{c}{Angular
size}&\colhead{}&\colhead{Critical redshift}\\
\tableline
 \colhead{}&\colhead{}&\colhead{$w$}&\colhead{$\chi^2_{min}$}&\colhead{$z_T$}&\colhead{$q_o$}&\colhead{}&\colhead{$w$}&
 \colhead{D}&\colhead{$\chi^2_{min}$}
 &\colhead{$z_T$}&\colhead{$q_o$}&\colhead{}&\colhead{$z_m$}\\
 \colhead{}&\colhead{}&\colhead{}&\colhead{}&\colhead{}&\colhead{}&\colhead{}&\colhead{}&
 \colhead{(mas)}&\colhead{}&\colhead{}&\colhead{}&\colhead{}&\colhead{}} \startdata
 1&&-0.7&15.4&-&-0.2&\phm{ih}&-0.45&1.25&4.72&-&0.05&&1.65\\
  &&-0.5&19.67&-&0&&-0.50&1.30&4.75&-&0&&1.72\\
  2&&1.1&16.5&0.45&-0.5&&0.50&1.45&4.47&1&-0.50&&1.71\\
  3&&1.7&16.35&0.42&-0.5&&1.15&1.40&4.58&0.77&-0.50&&1.65
\enddata
\end{deluxetable}

\clearpage

\begin{figure}[t]
\begin{center}
\includegraphics[scale=0.8]{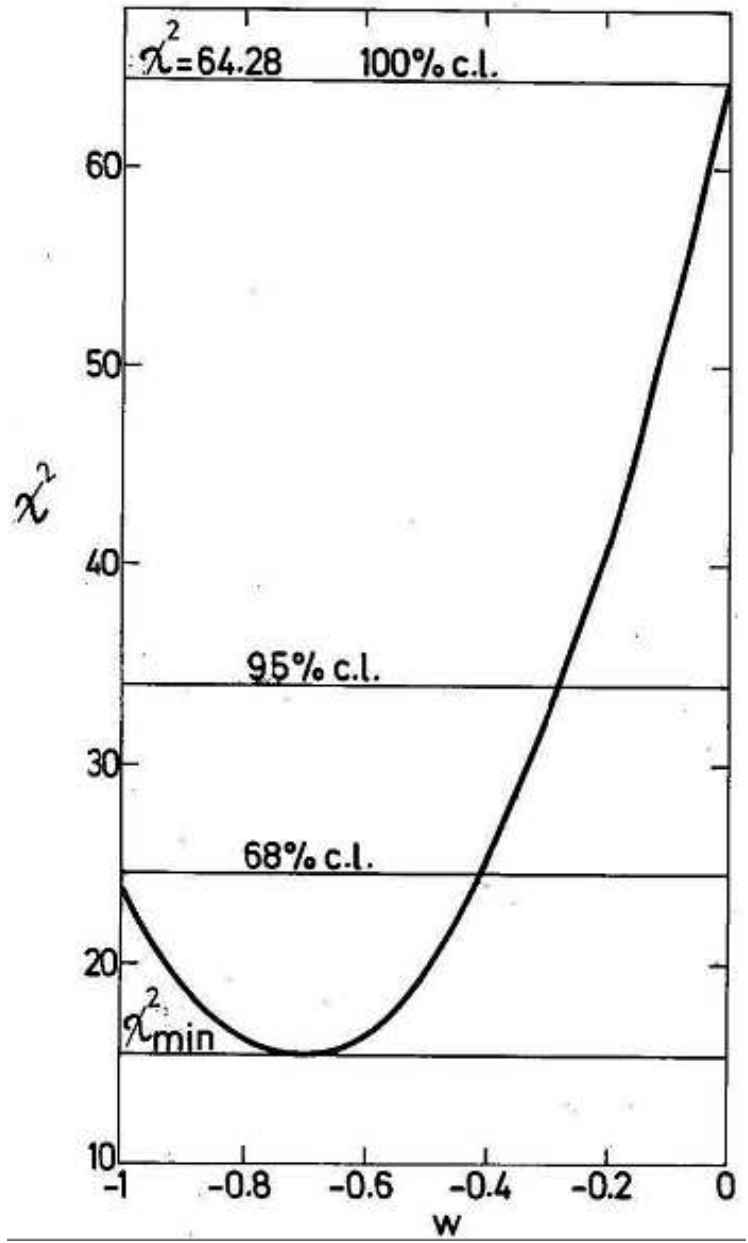}
\figcaption{Model 1 ($q \neq 0$): Plot of equation
(\ref{e21}), $\chi^2 $ versus $w$.}
\end{center}
\end{figure}
\clearpage
\begin{figure}[t]
\begin{center}
\includegraphics[scale=0.9]{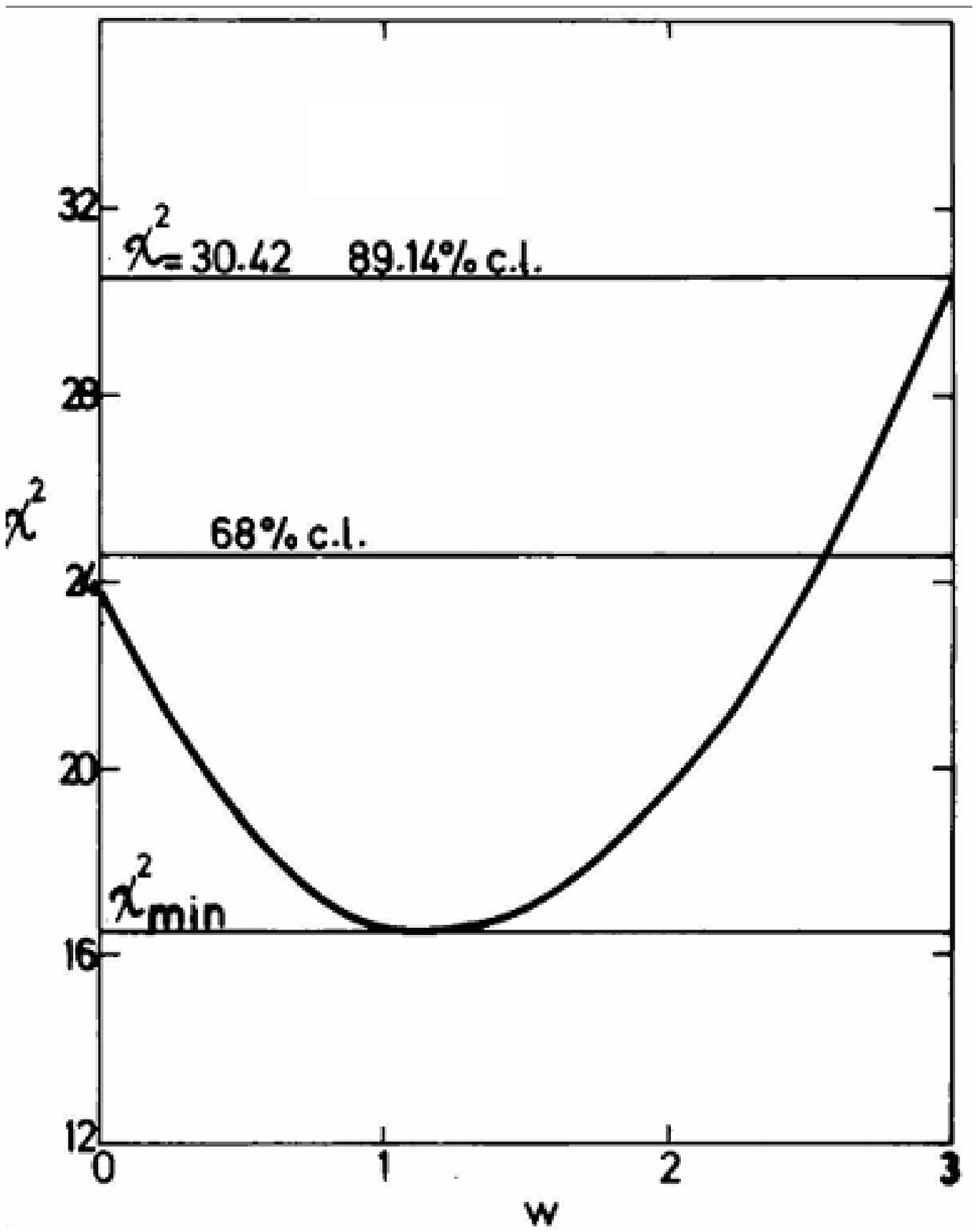}
 \figcaption{Model 2: Plot of equation
(\ref{e21}),$\chi^2 $ versus $w$.}
\end{center}
\end{figure}
\clearpage
\begin{figure}[h]
\begin{center}
\includegraphics[scale=0.9]{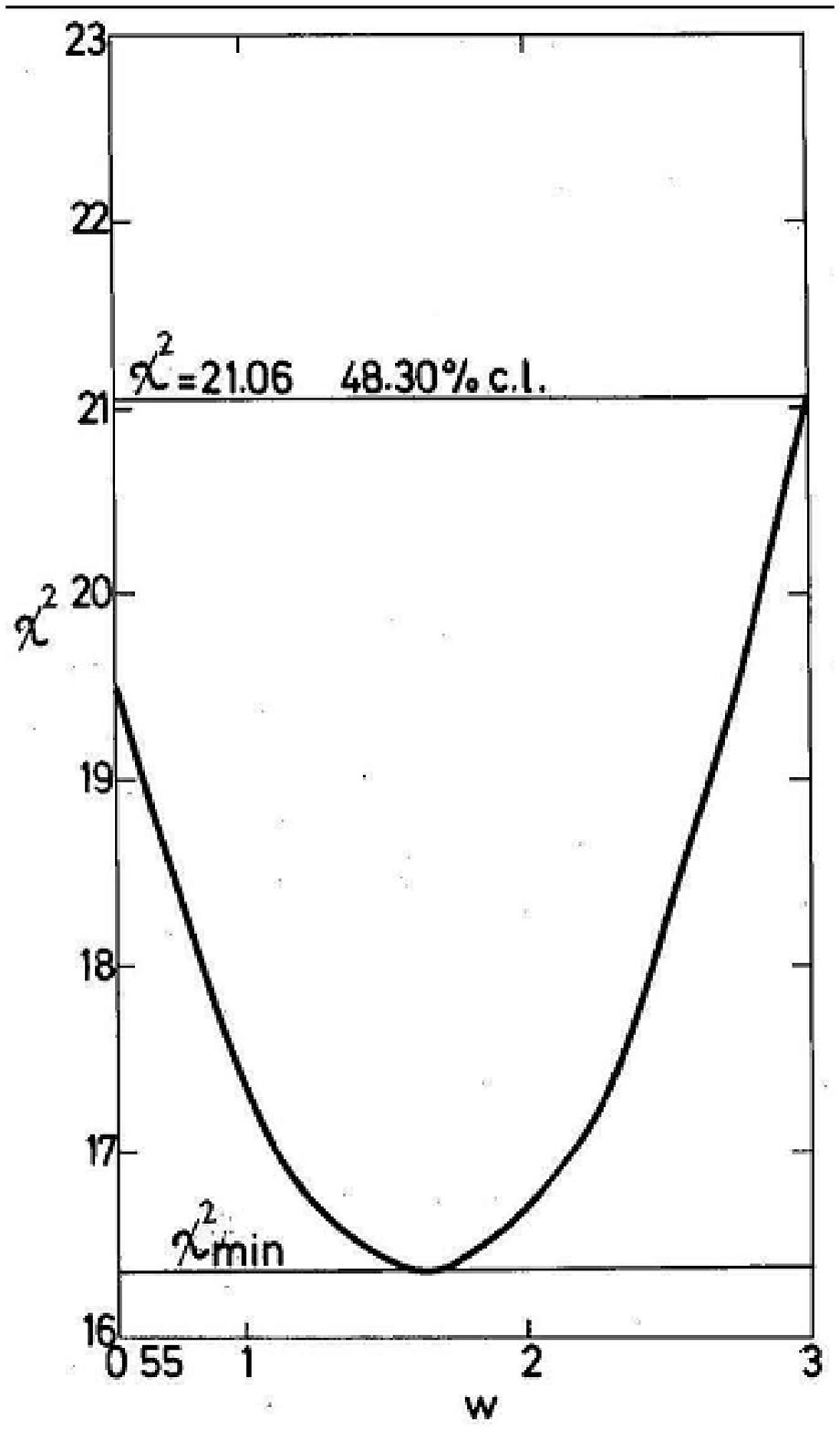}
\figcaption{Model 3:Plot of equation (\ref{e21}),$\chi^2 $
versus $w$.}
\end{center}
\end{figure}
\clearpage
\begin{figure}[h]
\begin{center}
\includegraphics[scale=0.9]{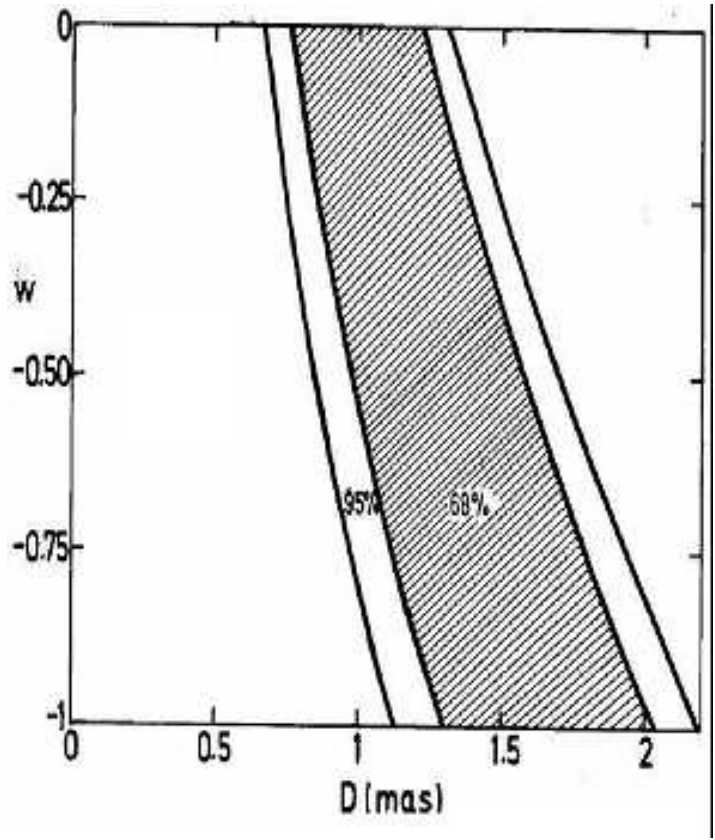}
\figcaption{Model 1 ($q \neq 0$) : Confidence contours in the
$ w - D $ plane.}
\end{center}
\end{figure}
\clearpage
\begin{figure}[h]
\begin{center}
\includegraphics[scale=1]{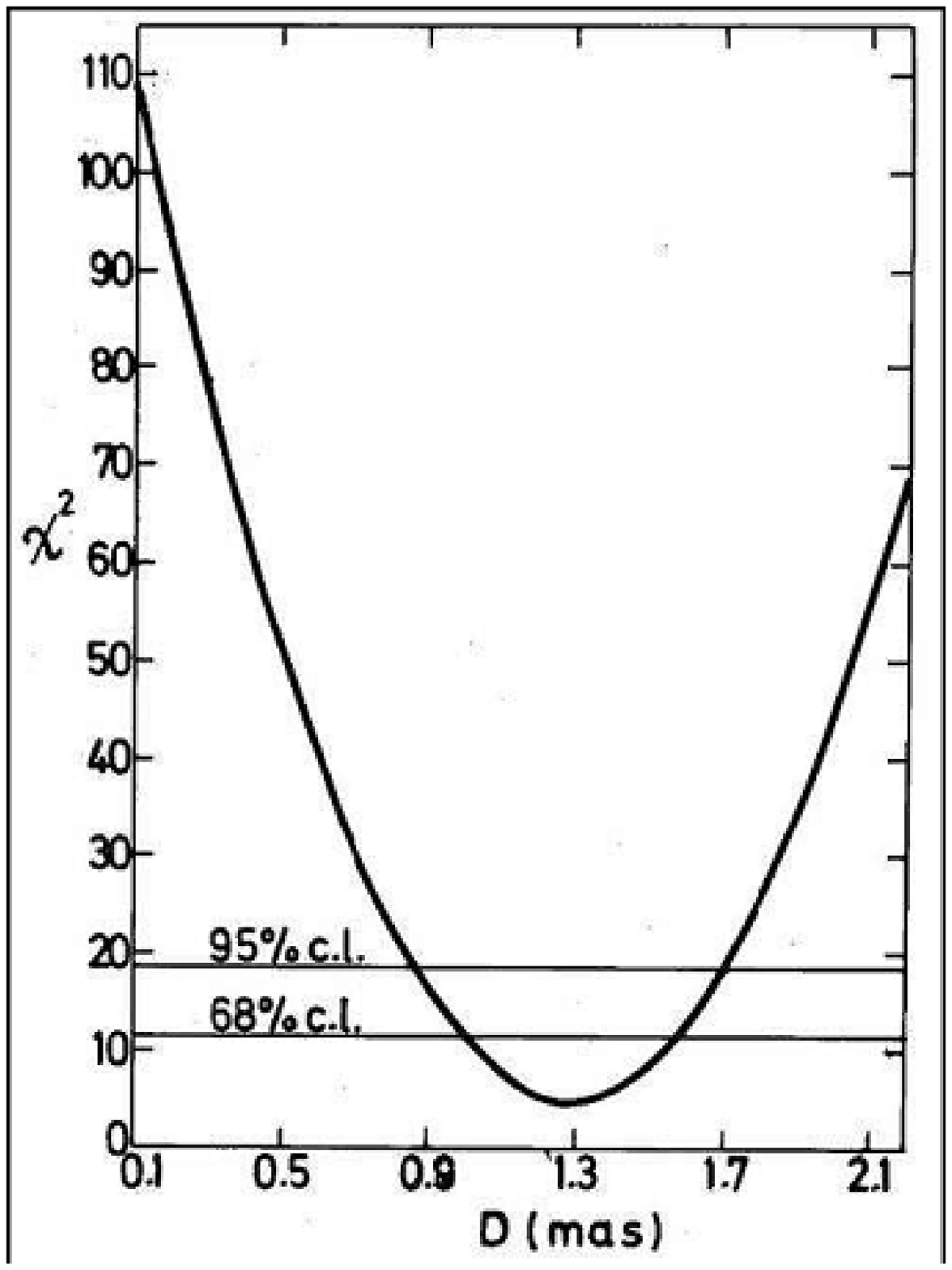}
 \figcaption{Model 1 ($ q = 0$): Plot of equation (\ref{e37}),
   $\chi^2 $ versus $D$.}
\end{center}
\end{figure}
\clearpage
\begin{figure}[h]
\begin{center}
\includegraphics[scale=0.9]{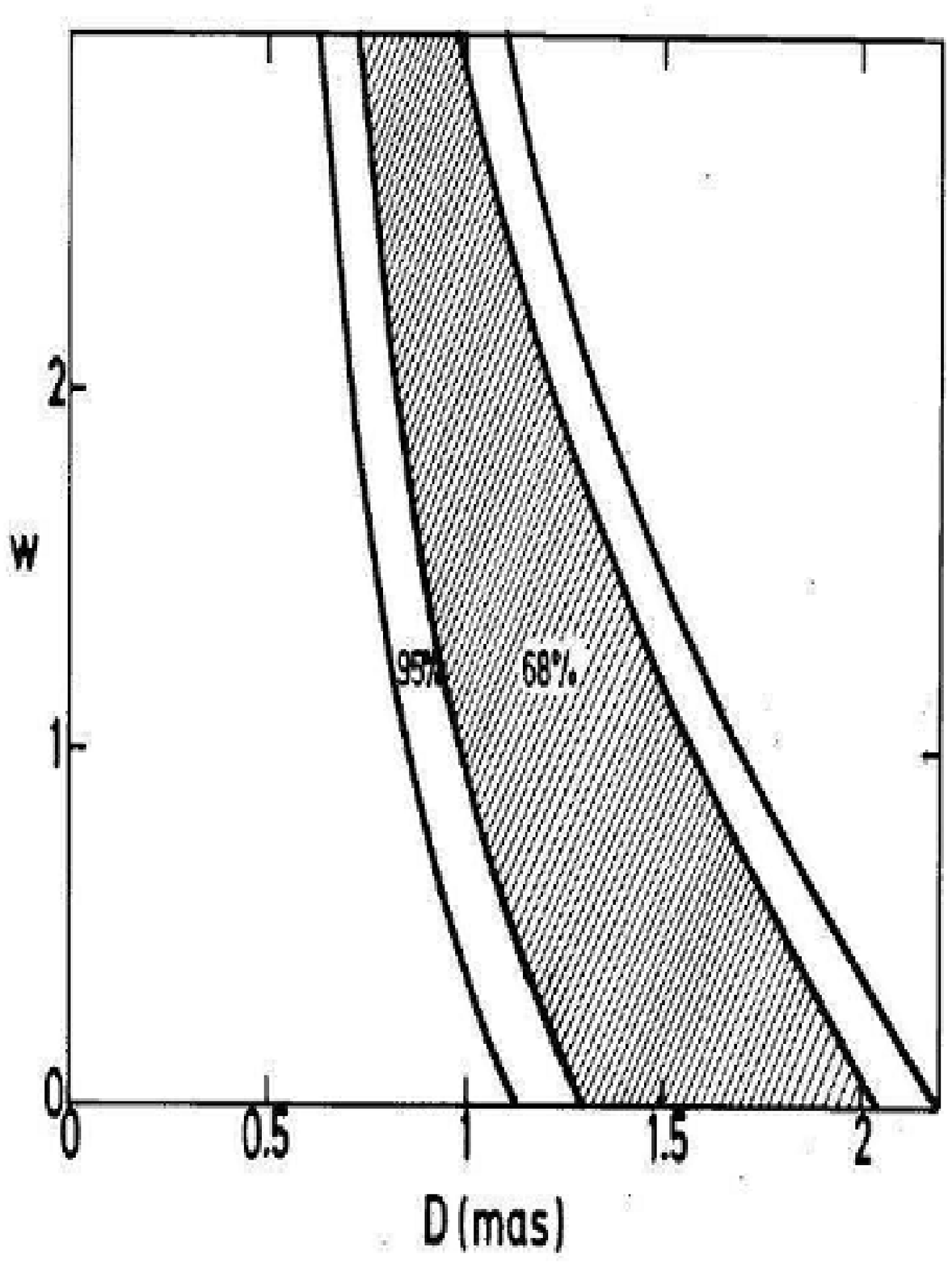}
\figcaption{Model 2: Confidence contours in the $w-D$
plane.}
\end{center}
\end{figure}
\clearpage
\begin{figure}[h]
\begin{center}
\includegraphics[scale=0.9]{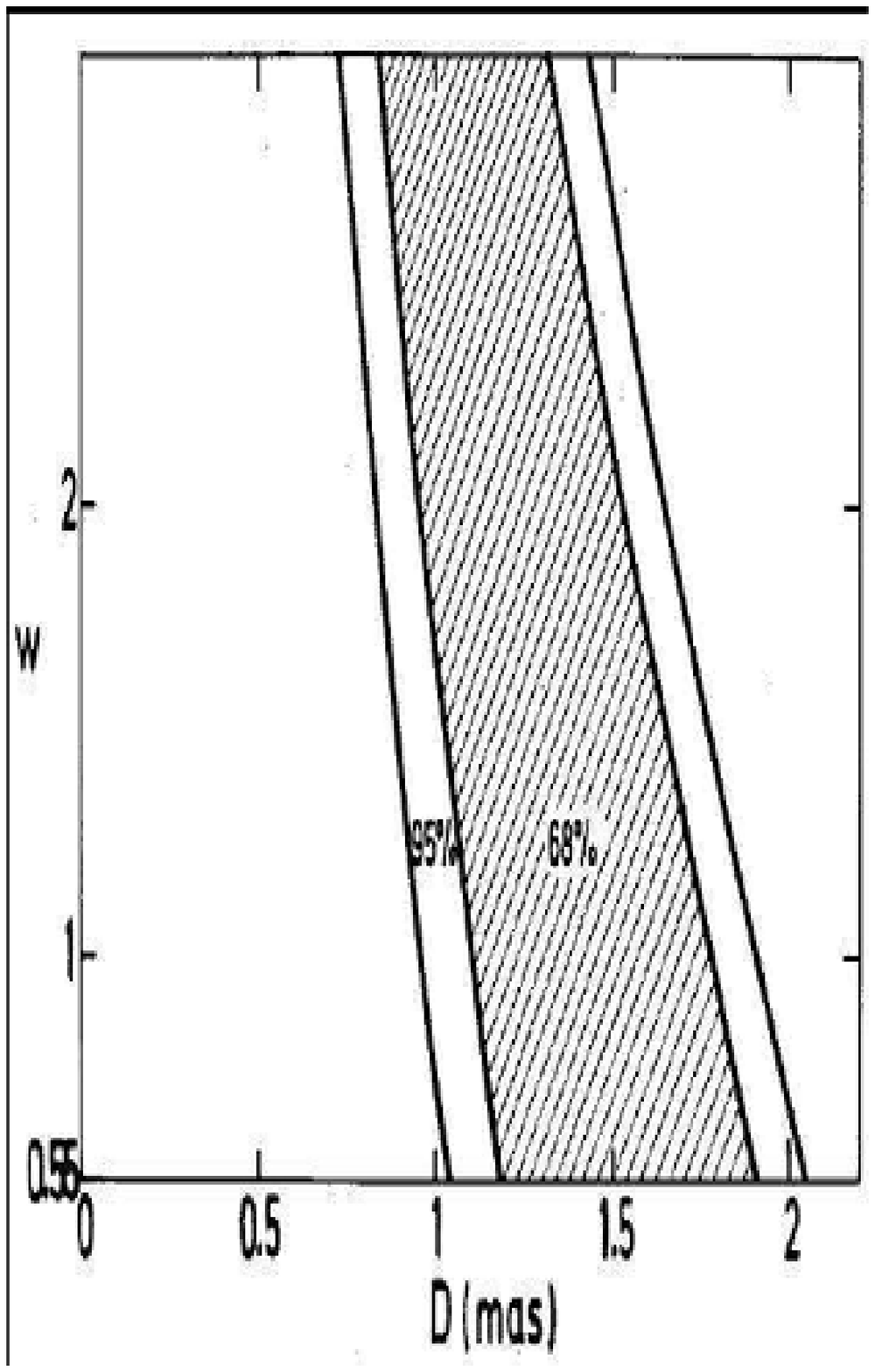}
\figcaption{Model 3: Confidence contours in the $w-D$
plane.}
\end{center}
\end{figure}
\clearpage
\begin{figure}[h]
\begin{center}
\includegraphics[scale=0.8]{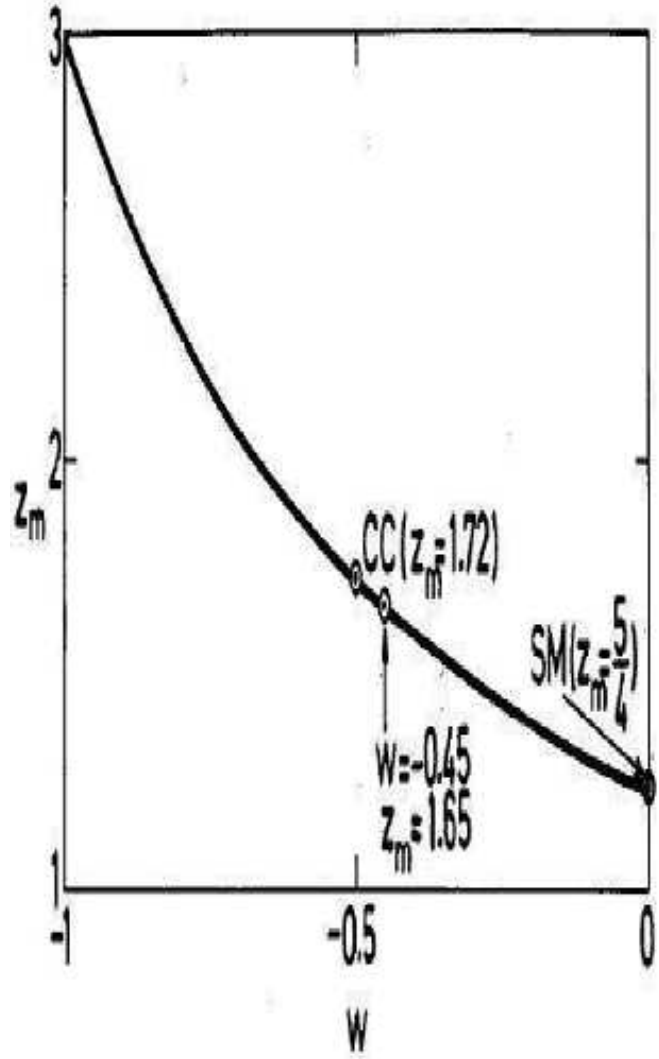}
\figcaption{Model 1 ($q \neq 0$) : Critical redshift versus
$w$.
    Also shown is the value $z_m = 1.72$ for the critical redshift in
     coasting cosmology (CC).}
\end{center}
\end{figure}
\clearpage
\begin{figure}[h]
\begin{center}
\includegraphics[scale=0.8]{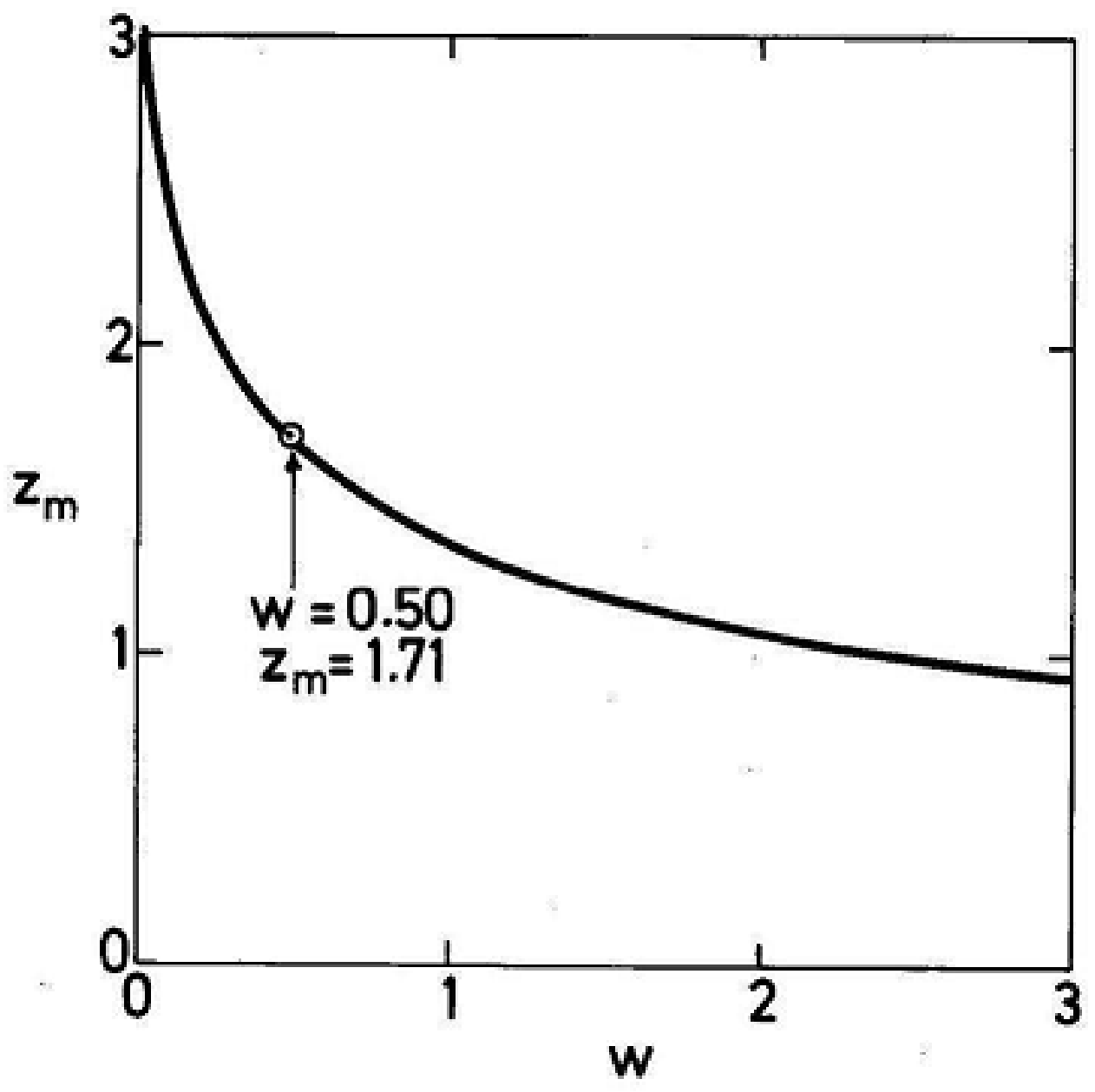}
 \figcaption{Model 2: Critical redshift versus $w$.}
\end{center}
\end{figure}
\clearpage
\begin{figure}[h]
\begin{center}
\includegraphics[scale=0.8]{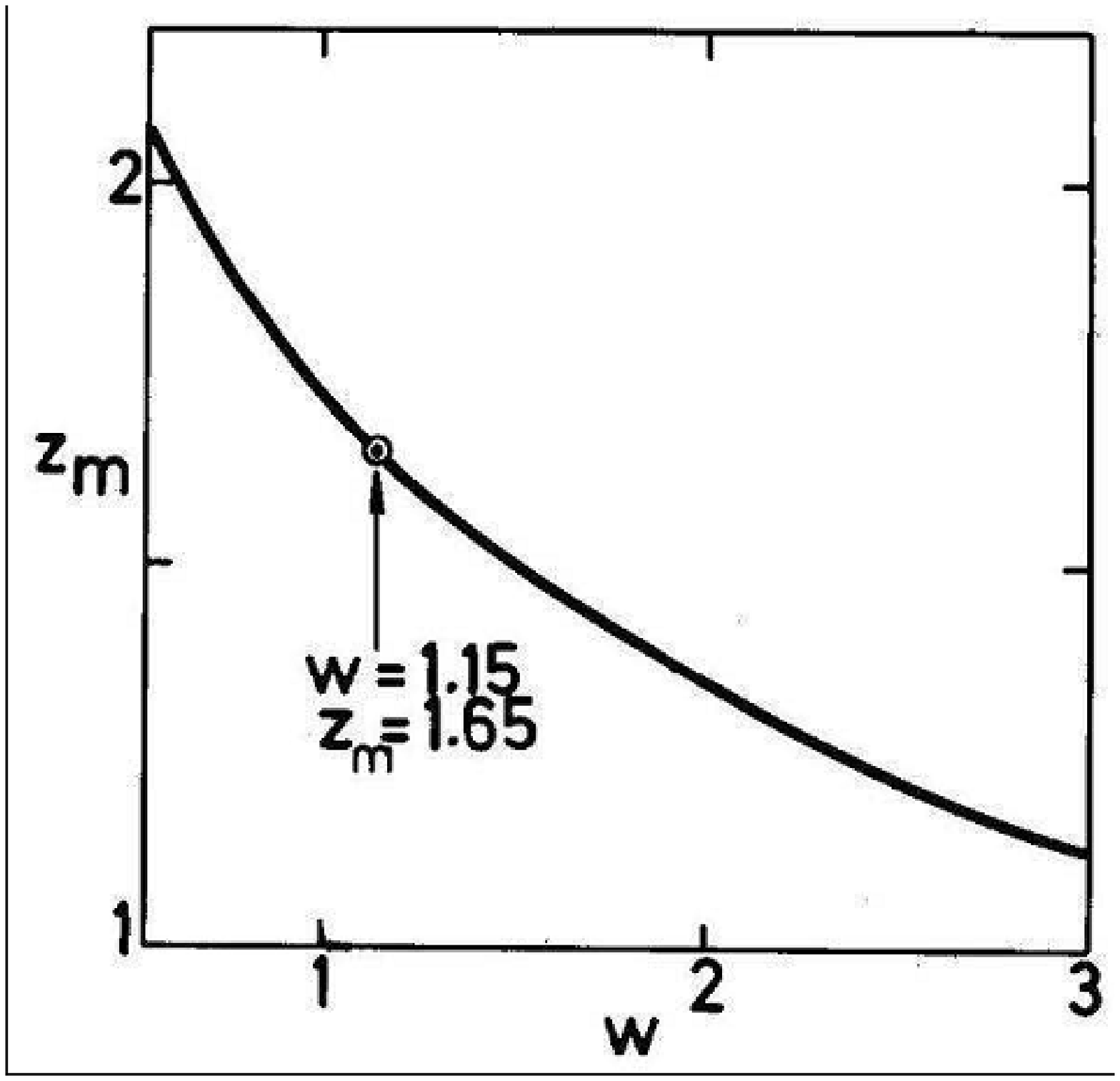}
 \figcaption{Model 3: Critical redshift versus $w$.}
\end{center}
\end{figure}
\clearpage
\begin{figure}[h]
\begin{center}
\includegraphics[scale=0.7]{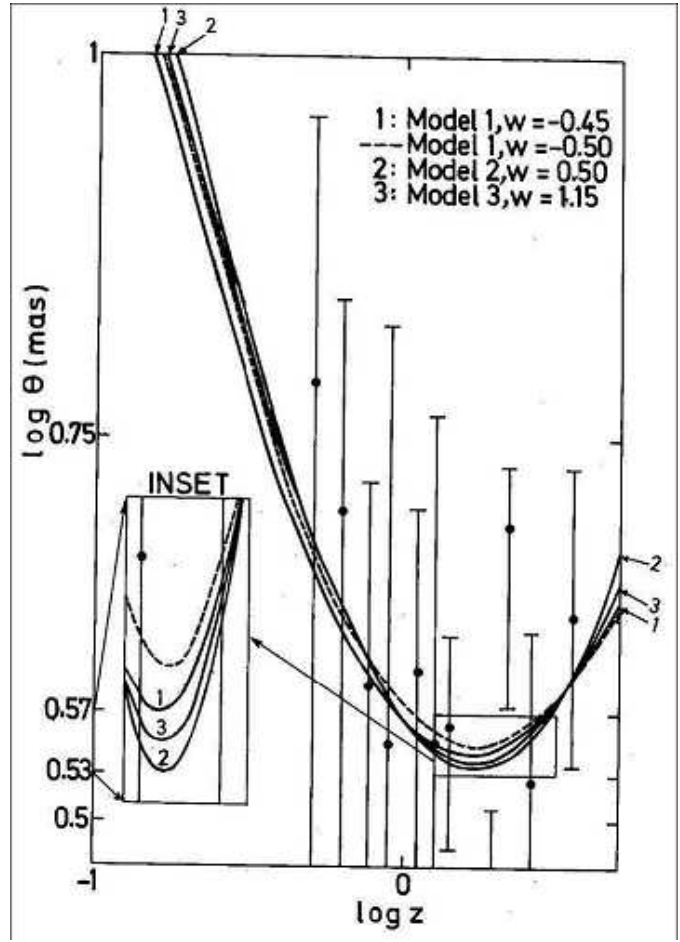}
\figcaption{Predicted $\theta - z $ curves for the models.
Also
   shown are the data points and error bars from
   \citet{GUR1999}.}
\end{center}
\end{figure}
\end{document}